# Superstructures par agrégation contrôlée de nanocolloïdes: caractérisation structurale par diffusion de neutrons aux petits angles et simulation numérique

Jean-François Berret[1] et Julian Oberdisse[2]

[1] *Université Paris-Diderot - Batiment Condorcet, Laboratoire Matière et Systèmes Complexes, UMR CNRS , 7057, 10 rue Alice Domon et Léonie Duquet, F-75205 Paris Cedex 13*

[2] *Université Montpellier II, Laboratoire des Colloïdes, Verres, et Nanomatériaux, UMR CNRS 5587, Place E. Bataillon, F-34095 Montpellier Cedex*

**Résumé.** La complexation de micelles ou de nanoparticules chargées par des copolymères à bloc chargés-neutres en solution aqueuse mène à la formation de superstructures colloïdales appelées aussi 'complexes de colloïdes'. Leur intérêt principal réside dans leur monodispersité en taille, ainsi que dans leur stabilité étendue. Dans cet article, nous présentons la caractérisation structurale de ces systèmes par diffusion dynamique de la lumière, cryo-TEM, et diffusion de neutrons aux petits angles. Les résultats de la diffusion aux petits angles ont été analysés à l'aide de simulations numériques – Monte Carlo ou Monte Carlo inverse (RMC) – prouvant la compatibilité entre différents modèles d'assemblages de colloïdes, et les expériences. Ces modèles ont donné accès aux paramètres structuraux comme le nombre de micelles ou de nanoparticules par superstructure, ou l'épaisseur de la couche de polymère pré-adsorbé sur les nanoparticules. L'ensemble des résultats nous a permis de proposer une structure générique, formée d'un cœur dense de colloïdes en interaction répulsive, pontés par les blocs de polyélectrolyte, et entouré d'une couronne très hydratée. Nous avons pu montrer que ces structures se forment systématiquement dans ces systèmes, qu'ils soient basés sur des micelles ou des nanoparticules, pour des copolymères variés, et de différentes charges.

## 1. Introduction générale

La compréhension de l'auto-assemblage de molécules de tensioactif en solution aqueuse a fait des grands progrès depuis les années 1980, notamment grâce aux expériences de diffusion de neutrons aux petits angles pour l'analyse structurale à l'échelle du nanomètre, comme évoqué dans le chapitre introductif de ce livre. D'un point de vue de la technique, la formation de micelles de tensioactifs – qui sont par nature presque parfaitement monodisperses - a considérablement facilité l'analyse quantitative, à cause du lissage moindre des courbes de diffusion par les effets de polydispersité. Ces dernières années, nous avons utilisé ces objets nanométriques bien définis comme briques de base pour la construction de superstructures. L'objectif de ce chapitre est d'en présenter l'analyse structurale détaillée, en combinant deux techniques de diffusion de rayonnement, lumière et neutrons, avec la microscopie électronique à transmission.

La formation de superstructures à partir de micelles en solution nécessite l'introduction d'une interaction attractive entre micelles. Une manière simple est d'ajouter des polyélectrolytes à une solution de micelles chargées électrostatiquement, et de signe opposé. Par un phénomène lié à l'adsorption des polymères à la surface des micelles et conduisant à leur pontage physique accompagné par la libération des contre-ions, une interaction effective attractive force les micelles à se coller les unes aux autres.

Comme nous le verrons dans la partie 'diagramme de phases', cette interaction est catastrophique, dans la mesure où elle colle toutes les micelles ensemble, et cause donc une séparation de phases macroscopique, dite 'attractive'. Une manière élégante d'éviter cette séparation de phases est l'utilisation de copolymères à bloc, faits d'un bloc hydrophile neutre en plus du bloc polyélectrolyte. Nous présenterons dans ce chapitre les superstructures générées avec des micelles, mais aussi la généralisation aux superstructures de nanoparticules d'oxyde. Ces dernières ont des applications intéressantes – catalyse, applications biomédicales, .. -, parce qu'elles peuvent apporter des fonctionnalités supplémentaires par leurs propriétés catalytiques, magnétiques ou optiques [1-5] . En pratique, il est souvent difficile de contrôler la taille de leurs agrégats en solution, et de garder la stabilité colloïdale. Nous verrons que les superstructures, que nous appelons aussi complexes de colloïdes, ne posent plus ces problèmes: comme les nanoparticules utilisées comme briques de base, ils sont bien définis en taille, et possèdent des domaines de stabilité élargis.

Dans la dernière partie de ce chapitre, nous présentons deux approches de simulation numérique utilisées pour l'analyse quantitative des courbes d'intensité de nos superstructures, car malheureusement une version analytique de leur facteur de forme ne fait pas partie des recueils publiés [6]. Comme déjà évoqué dans le chapitre de L. Belloni, la simulation directe d'un ensemble de particules avec des interactions supposées permet - en comparant la prédiction aux mesures de diffusion de neutrons aux petits angles - de valider les hypothèses sur ces interactions [7]. Nous avons appliqué cette idée aux complexes de micelles. Une autre méthode, dite de la simulation de Monte Carlo inverse, permet la génération d'une séquence de configurations probables d'un ensemble de nanoparticules, en accord avec l'expérience, ce qui revient à inverser de manière stochastique la transformation de Fourier faite lors de la mesure. Des propriétés statistiques peuvent ensuite être déduites des ensembles générés ainsi.

## 2. Matériaux et méthodes

Dans la partie consacrée aux superstructures de tensioactifs, nous illustrerons la technique de co-assemblage électrostatique à partir du système ternaire composé de copolymères à blocs chargés négativement, de tensioactif cationique et d'eau. Le polymère étudié est le poly(acide acrylique)-*b*-

poly(acrylamide) de poids moléculaire 6500 g mol$^{-1}$ et 37000 g mol$^{-1}$ pour chacun des blocs. Ce polymère est abrégé par PANa$_{6.5K}$-$b$-PAM$_{37K}$ dans la suite [8] Ces polymères ont été synthétisés par Rhodia en utilisant la technologie Madix en vue d'applications pour les formulations de produits de « Personal Care » [9]. Le tensioactif cationique est le bromure de dodecyltrimethylammonium (DTAB), qui est caractérisé par une chaîne aliphatique de douze atomes de carbone. Le système DTAB/eau forme une mésophase hexagonale à hautes concentrations [10]. Sa concentration micellaire critique (cmc) dans l'eau est de 0.46 wt. %, correspondant à 15 mM [11].

Dans la partie consacrée aux superstructures organique-inorganiques, nous utilisons des nanoparticules d'oxyde de fer de densité massique 5100 kg m$^{-3}$ chargées négativement, et des copolymères cationiques-neutres. Les particules d'oxyde de fer sont synthétisées en suivant la technique de Massart par co-précipitation alcaline de sels de fer II et III, par oxydation de la magnetite (Fe$_3$O$_4$) en maghemite ($\gamma$-Fe$_2$O$_3$) et par un tri en taille basé sur des séparations de phases successives [12, 13]. En fin de synthèse, les particules sont dispersées dans l'eau à une concentration massique de l'ordre de 10 wt. % et à pH 1.8. A ce pH, les particules sont chargées positivement. Les répulsions électrostatiques assurent la stabilité colloïdale des dispersions sur des périodes de plusieurs années. Les dispersions de nanoparticules d'oxyde de fer sont caractérisées par microdiffraction électronique, magnétométrie, magnétophorèse et diffusion de lumière. L'échantillon étudié ici a une distribution en taille qui a pu être ajustée par une fonction log-normale de diamètre médian $D_0$ = 6.3 nm et de polydispersité $\sigma$ = 0.23 (cf. aussi Fig. 6b) [14]. Les particules ont été ensuite fonctionnalisées avec des ligands citrate ou des oligomères de poly(acide acrylique), une procédure qui permet de modifier le signe des charges surfaciques grâce à l'ionisation des groupements carboxylates, de positif à bas pH à négatif à pH neutre. Pour les particules citratées (Cit–$\gamma$-Fe$_2$O$_3$), la densité de charges structurales a été estimée à -2$e$ nm$^{-2}$ par conductivité [15, 16] et par diffusion de la lumière [17]. Les particules Cit–$\gamma$-Fe$_2$O$_3$ ont donc des tailles et des densités de charges de surface supérieures à celles des tensioactifs présentés précédemment : 6.3 *versus* 4 nm pour le diamètre, -2$e$ *versus* +1$e$ nm$^{-2}$ pour la densité.

Le second système inorganique étudié dans cette revue est une dispersion de nanoparticules d'oxyde de cérium (CeO$_2$, ou nanoceria), synthétisée au Centre de Recherche d'Aubervilliers, Rhodia (France). La microscopie haute résolution électronique à transmission a montré que les nanoceria consistent en des agglomérats isotropes de cristallites de taille typique 2 nm et de morphologie facétée [18, 19]. Comme pour les particules de $\gamma$-Fe$_2$O$_3$, la distribution des tailles a été trouvée de type log-normale, avec une valeur médiane de 6.9 nm et une polydispersité de 0.15. Les particules, à l'origine cationiques, ont été ensuite recouvertes avec des ligands citrate, donnant des densités de charges électrostatiques comparables à celles des oxydes de fer [17]. Dans la suite, ces particules seront appelées Cit–CeO$_2$.

Parce que chargées négativement, les particules inorganiques d'oxyde de fer et de cérium ont été complexées par un copolymère séquencé de type cationique-neutre. Ce polymère, à nouveau synthétisé par l'équipe de Mathias Destarac à Rhodia, est le poly(trimethylammonium ethylacrylate methyl sulfate)-*b*-poly(acrylamide). Les poids moléculaires spécifiques à cette étude sont 5000 g mol$^{-1}$ et 11000 g mol$^{-1}$ pour la partie chargée et 30000 g mol$^{-1}$ pour la partie neutre. Nous appelerons ces polymères PTEA$_{5K}$-*b*-PAM$_{30K}$ et PTEA$_{11K}$-*b*-PAM$_{30K}$, respectivement. Dans un solvent aqueux, ces chaines sont à l'état d'unimères, donc non associées, avec des diamètres hydrodynamiques de l'ordre de 10 nm.

Le mélange des espèces de charges opposées a été fait en versant rapidement la solution de tensioactifs ou celle de nanoparticules dans la solution de polymères, au rapport de charge désiré. Ce protocole, nous l'avons appelé *mélange direct*, par opposition à d'autres protocoles développés ultérieurement tels que la *dialyse* ou la *dilution* [20-22]. Le rapport de charge a été calculé comme le rapport entre les charges structurales ionisables à la surface des micelles de tensioactifs ou des nanoparticules, et celles localisées le long des chaînes de polymères. Dans le reste de cette revue, les dispersions seront donc caractérisées par leur concentration totale c, c'est-à-dire la somme des concentrations des espèces, et le rapport de charge Z. Correspondant au mélange stoichiométrique de charges, les dispersions à Z = 1 ou Z voisins de l'unité seront en général celles sur lesquelles nous mettrons l'accent.

### 3. Comportement de phase

Comme annoncé dans l'introduction, il est possible d'induire des interactions attractives entre micelles chargées dans l'eau en ajoutant des polyélectrolytes de charge opposée. Dans un premier temps, nous avons étudié le diagramme de phases d'un mélange ternaire composé de DTAB, PANa, et d'eau. Sur la Figure 1a, on voit que le système présente une séparation de phases, pour les concentrations intermédiaires, et à partir d'un rapport de charge critique $Z_C > 0.1$. Rappelons qu'en augmentant Z, à concentration constante, on augmente la concentration en tensioactifs. A Z proche de zéro, la solution est essentiellement une solution de polymère, et quand Z tend vers l'∞, il s'agit d'une solution de tensioactifs. Dans la partie grisée de la figure 1a, nous avons trouvé pour c > 4 wt. % une séparation liquide-liquide, alors que pour c < 4 wt. % la séparation se fait entre une phase liquide et une phase solide. Les phases denses sont composées essentiellement des tensioactifs sous forme de micelles et des polymères, comme l'ont révélé des expériences de diffusion des rayons x [23]. Pour la phase solide apparaissant à basse concentration, nous avons confirmé qu'il s'agissait d'une mésophase crystalline de symétrie cubique (*Pm3n*) [24-27]. Notre interprétation de la séparation de phases est que les micelles, en forte interaction attractive grâce aux polyélectrolytes adsorbés à leur surface, forment un précipité en se séparant d'une phase aqueuse très diluée.

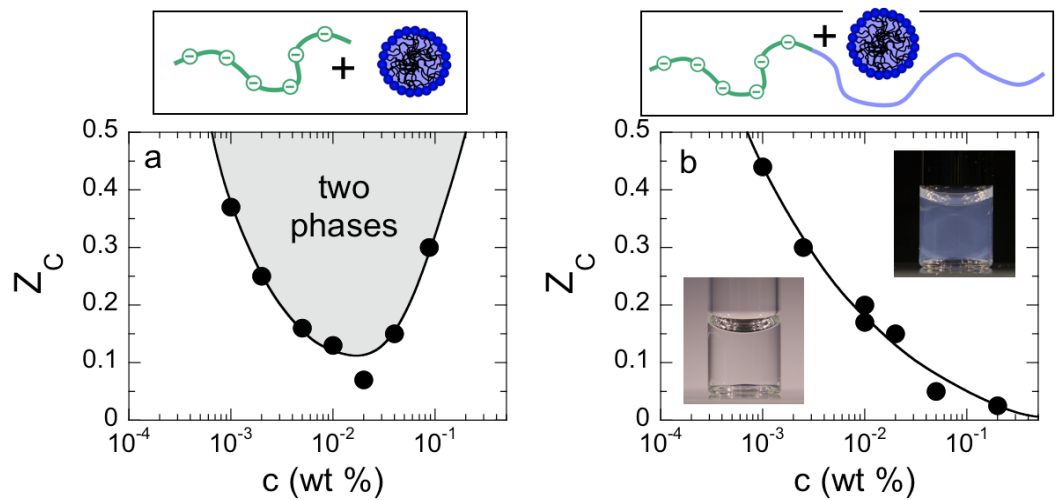

**Figure 1 :** **(a)** Diagramme de phases d'un mélange de polyélectrolyte - polyacrylate de sodium (PANa) - et de micelles de tensioactifs de charge opposée formées de DTAB, en fonction de la concentration c et du rapport des charges Z. Les conditions expérimentales sont pH 8 et T = 25°C. Le poids moléculaire du poly-anion sous sa forme acide est 5000 g mol$^{-1}$. **(b)** Le même diagramme de phases avec un copolymère PANa$_{6.5k}$-*b*-PAM$_{37k}$ à la place du polyélectrolyte.

Lorsque l'on construit le diagramme de phases dans le même espace de paramètres (c,Z) avec un copolymère doublement hydrophile, la séparation de phases disparaît. Rappelons que le copolymère est composé du même polyélectrolyte (PANa), avec un bloc hydrophile et neutre, le PAM (37k). Sur la Figure 1b, nous montrons le nouveau diagramme de phases, et à la place de la séparation de phases, nous observons deux régions monophasiques, diffusant la lumière différemment. Aux faibles concentrations, les échantillons sont transparents (cf. photo de gauche). Aux concentrations plus élevées, les échantillons deviennent diffusants (cf. photo à droite), et les deux zones sont séparées par une ligne Z(c) bien définie. Une diffusion élevée suggère la formation d'agrégats plus grands que les éléments constitutifs que sont les micelles et les polymères. Nous allons étudier la microstructure de ces solutions dans la section suivante.

Pour des raisons pratiques, il n'est pas possible de construire un diagramme analogue avec les particules inorganiques de tailles nanométriques à la place des micelles de tensioactifs, notamment à cause des concentrations difficiles à atteindre. Pour le système homopolymères/nanoparticules, nous avons toujours observé une séparation de phases sur les domaines qui sont ici représentés. Avec les copolymères diblocs, le diagramme de phases complet n'a pas été établi, mais on trouve une phase homogène semblable avec une forte diffusion indicatrice de la présence d'une organisation supra-colloïdale de la matière.

**4. Caractérisation structurale**

Dans cette section, nous présentons des résultats obtenus par diffusion dynamique de la lumière et diffusion de neutrons aux petits angles. Dans les deux techniques, on envoie un faisceau sur un échantillon, qui est une solution colloïdale dans notre cas. Le rayonnement diffusé est détecté. Pour les systèmes isotropes, la variable pertinente est la norme du vecteur de diffusion q, qui est la différence entre le vecteur d'onde incident et diffusé [28].

Comme la longueur d'onde de la lumière est proche du micron, on sonde des distances à cette échelle environ. Le phénomène de diffusion est induit par des fluctuations de densité dans la solution colloïdale à cette échelle. En suivant la diffusion dans le temps, on peut remonter au diamètre hydrodynamique $D_H$ des particules :

$$D = \frac{\Gamma}{q^2} = \frac{k_B T}{3\pi\eta D_H} \qquad (1)$$

Ici, la première égalité permettant la mesure du coefficient de diffusion D vient de l'équation de diffusion des particules, $\Gamma$ étant la constante de décroissance du champ observé en diffusion de la lumière au vecteur d'onde q. La deuxième est la relation de Stokes-Einstein. Elle fait intervenir l'énergie thermique $k_B T$, $k_B$ étant la constante de Boltzmann, et la viscosité du solvant $\eta$. L'équation (1) s'applique si le mouvement est diffusif (Brownien), par exemple s'il n'est pas dominé par des interactions à longue distance. En variant l'angle de diffusion et donc q, on peut vérifier la validité de l'équation de diffusion. Notons que même si la diffusion de la lumière est observée à l'échelle micronique, l'équation (1) permet de la relier à la diffusion Brownienne de particules beaucoup plus petites, comme par exemple les micelles.

Dans les expériences de diffusion de neutrons aux petits angles, on ne s'intéresse qu'à la moyenne de l'intensité diffusée, en fonction de l'angle de diffusion et donc du vecteur d'onde q. Comme la longueur d'onde des neutrons est de l'ordre de 0.5 à 1 nm, l'échelle sondée est beaucoup plus petite que celle de la lumière visible. Pour des expériences aux petits angles, l'inverse du vecteur d'onde, qui caractérise cette échelle, peut varier d'un à quelques dizaines, voire des centaines, de nanomètres. Typiquement, cette méthode nous donne donc accès à la structure et à l'agrégation de micelles ou nanoparticules, dont la taille typique est de l'ordre de quelques nanomètres.

### 4.1 Diffusion dynamique de la lumière

#### 4.1.1 Les superstructures à base de micelles vues par DLS

Nous avons vu en section trois que l'ajout de copolymères doublement hydrophiles à des solutions micellaires modifie radicalement le diagramme de phases. La séparation de phases attractive observée dans le cas des homopolymères est remplacée par une phase macroscopiquement homogène, mais présentant une forte diffusion de la lumière observable à l'œil nu pour des quantités de micelles importantes (grand Z).

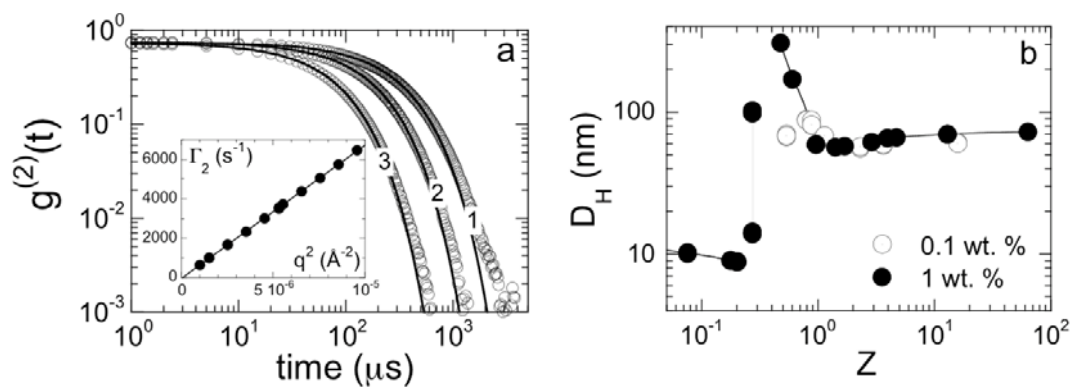

**Figure 2 :** Diffusion de la lumière du système DTAB/ PANa$_{6.5k}$-$b$-PAM$_{37k}$. **(a)** Trois exemples de fonctions de d'autocorrélation de l'intensité diffusée en fonction du temps de corrélation, pour différents vecteurs de diffusion q. La décroissance exponentielle est décrite par une constante $\Gamma$, dont la dépendance de $q^2$ est représentée dans l'insert. **(b)** Diamètre hydrodynamique des superstructures à concentration fixée (c=1 wt. %), en fonction du rapport de charge Z.

Il paraît donc logique de mesurer la diffusion de la lumière par DLS, en suivant une ligne en Z, dans le système formé par le DTAB et le copolymère PANa$_{6.5k}$-$b$-PAM$_{37k}$. Pour ces expériences, la diffusion statique et dynamique de la lumière a été obtenue en utilisant un spectromètre Brookhaven (BI-9000AT autocorrelator, $\lambda$ = 488 nm) et un NanoZS (Malvern instruments, $\lambda$ = 633 nm) pour la mesure du rapport de Rayleigh et du coefficient de diffusion D. Pour un Z donné, la mesure consiste à suivre I(q,t) en fonction du temps, et d'en déterminer la fonction de corrélation. En Figure 2a, la fonction de d'autocorrélation est montrée pour trois angles de diffusion différents, afin de vérifier la diffusivité

Brownienne des particules. En représentant la constante de décroissance $\Gamma$ en fonction de $q^2$, cf. l'insert de la Figure 2a, on se convainc facilement de la linéarité postulée en équation (1).

On peut donc en déduire le coefficient de diffusion D donné par la pente de la droite. En Figure 2b, nous montrons le résultat pour différents Z, rapporté en termes du diamètre hydrodynamique $D_H$. Dans les solutions contenant quasiment exclusivement du copolymère (Z proche de zéro), le diamètre hydrodynamique est faible. Vers Z = 0.2 - 0.3, la taille des nanostructures augmente considérablement, passe par un maximum, et se stabilise vers 80 nm environ pour les Z > 1. La valeur du rapport de charge à laquelle la transition a lieu nous a permis de tracer la ligne de séparation dans le diagramme de phases, Figure 1b. En résumé, la diffusion dynamique de la lumière nous apprend que les micelles et le copolymère forment des grands objets, de diamètre typique de 80 nm, pour des rapports de charge supérieurs à un rapport critique de l'ordre de un. Comme nous savons par ailleurs que la structure des micelles est préservée dans les précipités, il est probable que les grands objets visibles en diffusion de la lumière soient des superstructures de micelles. Par la suite, nous allons montrer qu'un phénomène analogue se produit avec les nanoparticules, avant d'étudier la structure interne des grands objets par diffusion de neutrons.

*4.1.2 Les superstructures à base de nanoparticules vues par DLS*

Nous allons maintenant vérifier que l'on peut former le même type de superstructures avec des nanoparticules de $CeO_2$ et de $\gamma$-$Fe_2O_3$. Le copolymère doublement hydrophile utilisé dans cette étude est le PTEA$_{5K}$-*b*-PAM$_{30K}$, i.e. c'est le même bloc neutre que précédemment, mais un polyélectrolyte différent [29], afin d'être de charge opposée par rapport aux nanoparticules qui sont elles négativement chargées en solution à pH 7.

Pour les deux types de nanoparticules, nous observons un plateau où le diamètre hydrodynamique reste constant, $D_H \approx 65$ à 70 nm, puis une décroissance vers la taille des nanoparticules individuelles pour Z très grand, i.e. dans des solutions contenant essentiellement des nanoparticules. Pour les valeurs de Z > 10, on observe en fait une coexistence de deux modes de diffusion : un mode rapide qui correspond aux nanoparticules individuelles et un mode lent qui est associé aux colloïdes de taille 70 nm. Cette coexistence à forts Z, indiquée par les traits verticaux a été interprétée en termes de complexation à rapport stoichiométrique entre particules et polymères constant [17].

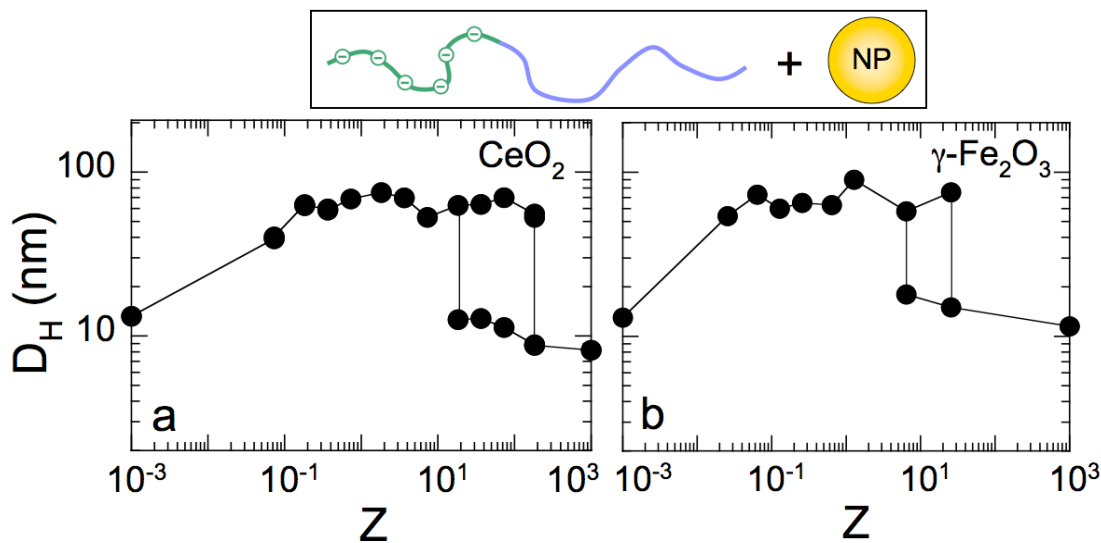

**Figure 3 :** **(a)** Diamètre hydrodynamique de superstructures de nanoparticules d'oxyde de cérium et de copolymère PTEA$_{5K}$-*b*-PAM$_{30K}$ en fonction du rapport de charge Z. **(b)** Même observation pour les nanoparticules d'oxyde de fer.

### 4.2  Microscopie électronique à transmission (TEM)

*4.2.1 Les superstructures à base de micelles vues par cryo-TEM*

Les observations par DLS de la taille hydrodynamique suggèrent que les petites micelles ou les nanoparticules forment des objets beaucoup plus grands lorsqu'une valeur critique de Z est franchie. Rappelons que le diamètre de ces grands objets est typiquement de l'ordre de 60 à 80 nm. Pour compléter l'analyse structurale à cette échelle, nous avons fait des expériences de microscopie électronique. Pour le système tensioactif-copolymère (DTAB/PANa$_{6.5k}$-*b*-PAM$_{37k}$), un cliché typique de cryo-TEM est montré en Figure 4. On y observe une multitude de petites taches, de diamètre environ 25nm. Notons aussi que ces objets sont plutôt monodisperses.

A ce stade de l'étude nous ne pouvons que constater que la taille typique des objets en suspension dans un mélange de micelles chargées et de copolymères chargé-neutres est plus grande que celle des micelles individuelles. Cependant, la DLS et la cryo-TEM nous fournissent deux tailles différentes, environ 70 et 25 nm : visiblement ces deux mesures ne caractérisent pas les mêmes objets, ou pas les mêmes dimensions des superstructures. Avant de nous intéresser à la structure sondée par la diffusion de neutrons aux petits angles, regardons d'abord si les superstructures de nanoparticules ont le même comportement particulier.

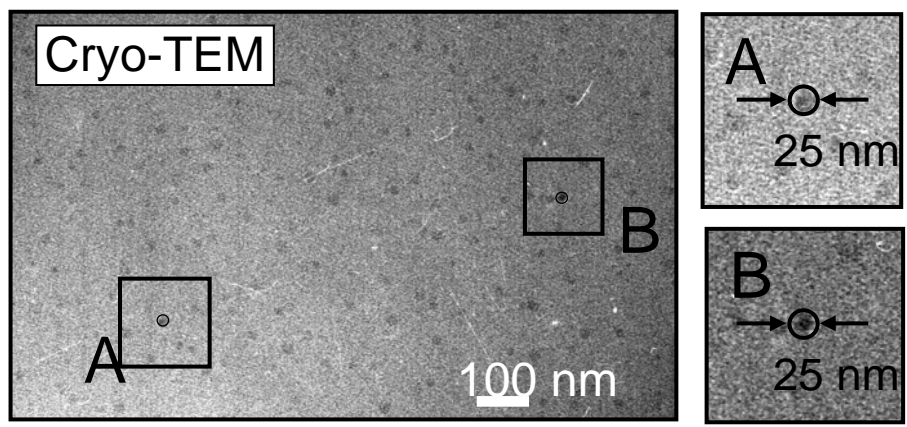

**Figure 4 :** Cliché de microscopie cryogénique à transmission (cryo-TEM) du système DTAB/PANa$_{6.5k}$-$b$-PAM$_{37k}$ (c = 1 wt. %, Z = 1).

*4.2.2 Les superstructures à base de nanoparticules vues par cryo-TEM*

Nous avons répété les observations par cryo-TEM avec les systèmes de nanoparticules et copolymères. Les résultats sont montrés en Figure 5 pour les oxydes de cérium et de fer. Grâce à un contraste amélioré par rapport aux tensioactifs (ce contraste est dû à la présence d'ions métalliques de cérium ou de fer de nombres atomiques élevés), on distingue bien les nanoparticules sur les clichés de microscopie électronique.

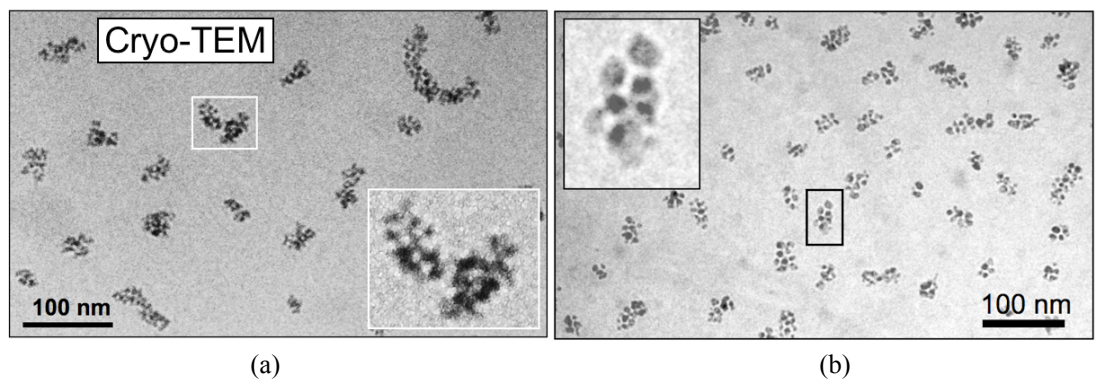

(a)        (b)

**Figure 5 :** Cliché de cryo-microscopie à transmission (cryo-TEM) des systèmes nanoparticules-PTEA$_{5K}$-$b$-PAM$_{30K}$. **(a)** Nanoparticules de CeO$_2$ (diamètre D$_0$ = 6.9 nm, c= 0.2 wt. %). **(b)** Nanoparticules de $\gamma$-Fe$_2$O$_3$ couvertes de PAA (D$_0$ = 6.3 nm, c= 0.2 wt. %). Les solutions ont été préparées par mélange direct au rapport stoichiométrique (Z = 1).

Ces nanoparticules forment des grandes superstructures, de taille relativement monodisperse. Nous n'observons ni agrégats géants, ni particules individuelles. Le rapport d'aspect des agrégats est en général légèrement supérieur à un (objets allongés). La taille globale est de l'ordre de 25 nm, de nouveau notablement plus petite que la taille hydrodynamique, qui était de 70 nm environ.

En vue des résultats concordants pour les systèmes de micelles et de nanoparticules, deux questions se posent : (a) quelle est la structure interne des grands objets, et comment réconcilier les deux tailles observées en DLS et TEM? (b) est-ce que le tensioactif est toujours sous forme de micelles dans les objets plus grands ? Nous avons déjà apporté un élément de réponse à la deuxième question, dans la mesure où nous savons que les micelles existent toujours au sein du précipité formé par les polyélectrolytes [23]. Dans le cas des nanoparticules, elles existent clairement toujours, comme on le voit très bien en Figure 5. Dans la suite de ce chapitre, nous avons cherché à caractériser la structure interne des superstructures par la diffusion de neutrons aux petits angles afin de répondre à la première question.

### 4.3 Diffusion de neutrons aux petits angles

Nous avons fait des expériences de DNPA afin de caractériser la structure de nos colloïdes et superstructures colloïdales en solution. Les expériences ont été faites sur trois sites, Argonne National Laboratory, USA; Laboratoire Léon Brillouin et Institut Laue-Langevin, France. Tous les résultats étaient cohérents d'un spectromètre à l'autre. Ici, nous montrerons des données obtenues sur les lignes D11 et D22 de l'Institut Laue-Langevin. Les solutions ternaires de tensioactifs et copolymères étaient préparées à la concentration typique c = 1 wt. % en utilisant $D_2O$ comme solvant, pour des raisons de contraste. Pour les mélanges de particules nanométriques inorganiques, le solvant était $H_2O$, et cela pour deux raisons : les synthèses chimiques étaient faites dans l'eau légère, et le contraste en neutron pour ces particules est meilleur dans ce solvant.

Comme cela a été écrit dans les chapitres précédents, nous avons mesuré la section efficace différentielle par unité de volume d'échantillon $d\Sigma/d\Omega$, et plus exactement sa partie cohérente, donnée en $cm^{-1}$. Pour faire court, nous appelons cette fonction de diffusion 'intensité', ou $I(q)$. Sa dépendance en q nous donne des informations sur la forme des superstructures. Dans le cas idéal, nous observons:

(i) en absence d'interactions, i.e. à dilution suffisante, un plateau aux petits angles, $I_0 = I(q \rightarrow 0)$. Ce plateau nous renseigne sur la masse totale des espèces visibles dans une superstructure. Pour un seul type d'objets, de contraste $\Delta\rho$ homogène, $I_o$ est donnée par:

$$I_0 = \Phi \, \Delta\rho^2 \, V_{tot} \qquad (2)$$

où $\Phi$ est la fraction volumique d'objets, et $V_{tot}$ le volume sec d'un objet. C'est ce terme qui nous permet de dire que l'on mesure la 'masse' aux petits angles, comme dans les expériences de diffusion de lumière statique.

**(ii)** une décroissance rapide avec q. Le tout début de cette décroissance nous permet d'estimer le rayon de giration $R_g$, grâce à l'équation de Guinier:

$$I(q) = I_0 \exp\left(-\frac{q^2 R_G^2}{3}\right) \tag{3}$$

**(iii)** Plus loin, aux q intermédiaires, on trouve parfois des oscillations, qui traduisent une certaine monodispersité des superstructures.

**(iv)** Aux grands q, on observe l'intérieur des complexes, i.e. les micelles ou nanoparticules constituant le cœur, en interaction.

Dans la discussion ci-dessus, nous avons caché une petite subtilité dans le mot 'visible'. Toutes les molécules et particules appartenant à une superstructure participent à la diffusion, si elles possèdent un contraste avec le solvant. Dans le cas de micelles, nos expériences ayant été faites dans du $D_2O$, les deux constituants, polymère et micelles, devraient être normalement visibles. Cependant, le PAM a la particularité d'échanger ses protons avec des D provenant de l'eau lourde, et son contraste effectif devient très faible [30]. Dans le cas des nanoparticules, la question ne se pose pas, les mesures étant en général faites dans de l'eau légère. Par ailleurs, la diffusion par des particules denses domine en général celle provenant des polymères. Pour ces deux raisons, nous pouvons nous contenter de discuter la diffusion par les micelles ou nanoparticules uniquement. La courbe décrivant le facteur de forme des micelles ou nanoparticules seules, i.e. en solution diluée, sans polymère, se présente schématiquement de la même manière que décrite ci-dessus : plateau, décroissance, structure à petite échelle (surface). La masse qui détermine le niveau de diffusion aux petits angles, $I_o$, est plus faible, et – en réduisant à des conditions de contraste et concentration identiques -, le rapport de $I_o$(superstructure) et $I_o$(nanoparticule) permet d'extraire le nombre d'agrégation moyen, $N_{agg}$ :

$$N_{agg} = \frac{\dfrac{I_{0,s}}{\Phi_s \Delta\rho_s^2}}{\dfrac{I_{0,NP}}{\Phi_{NP} \Delta\rho_{NP}^2}} \tag{4}$$

Nous présentons par la suite les résultats de nos expériences de DNPA, avec les superstructures formées de micelles et de nanoparticules.

*4.3.1 Les superstructures à base de micelles vues par DNPA*

En Figure 6a, nous avons représenté un résultat-clé de nos campagnes de mesure de diffusion de neutrons aux petits angles. Toutes les intensités ont été mesurées à une concentration massique de 1 wt. %. Le cas limite, $Z = \infty$, présente bien une diffusion de micelles individuelles et petites, comme on le comprend tout de suite à la vue du niveau général faible de l'intensité. L'intensité possède un maximum juste avant 0.05 Å$^{-1}$, qui nous dit que les micelles interagissent par répulsion électrostatique, même à cette dilution. La diminution de l'intensité aux petits angles reflète la faible compressibilité osmotique, et elle pourrait être décrite par un facteur de structure de sphères chargées [7, 31]. Une telle description nous donnerait alors accès à la limite à angle nul du facteur de forme des micelles, $I_o(c \rightarrow 0, Z=\infty)$, et donc leur masse par l'éq. (2). Nous avons pu la mesurer en ajustant un facteur de forme sur la courbe expérimentale [8], et nous avons trouvé que les micelles de tensioactifs de DTAB ont un nombre d'agrégation de 53 ± 3 et un diamètre de 3.63 nm. Notons aussi que la diffusion aux grands angles correspond à la diffusion par la surface des micelles.

Dans le cas du polymère seul ($Z = 0$, données non montrées), la courbe n'a pas une allure très marquée, ce qui reflète la conformation typique d'une molécule de polymère en solvant. Le cas qui nous intéresse, $Z = 1$, c'est-à-dire un mélange stoichiométrique en charges de copolymère et de tensioactifs, possède une intensité beaucoup plus élevée que les deux cas limites. L'allure générale de la courbe est bien décrite par les points (i) à (iv) dans l'introduction de la section 4.3. L'existence d'un domaine de Guinier nous indique immédiatement qu'il s'agit d'objets de taille finie, et plutôt bien définie, comme nous pouvons déduire de l'existence d'une oscillation amortie aux moyens angles.

Si on ignore la faible contribution des copolymères à la diffusion, le rapport des intensités aux petits angles, $I_o(Z=1)/I_o(Z=\infty)$ nous permet d'estimer le nombre moyen de micelles dans un tel objet en suivant l'équation (4), et nous obtenons une centaine de micelles par superstructure. Une évaluation plus précise nous confirme cette valeur [8]. Une méthode simple pour amplifier l'oscillation amortie aux moyens angles est de la reproduire en représentation de Porod, $q^4 I(q)$ en fonction de q (ou parfois de $q^4$). Pour le mélange à $Z = 1$, cette courbe est montrée en Figure 6b. On observe de jolies oscillations amorties, qui rejoignent un plateau, dit plateau de Porod. Etant donné la multiplication par $q^4$, la hauteur de ce plateau n'est rien d'autre que le préfacteur de la décroissance de Porod aux grands angles, qui, elle, est due aux surfaces lisses des objets. Un ajustement avec des sphères polydisperses nous permet de reproduire les oscillations, et en déduire un diamètre moyen de 21.6 nm, avec une polydispersité de 0.16.

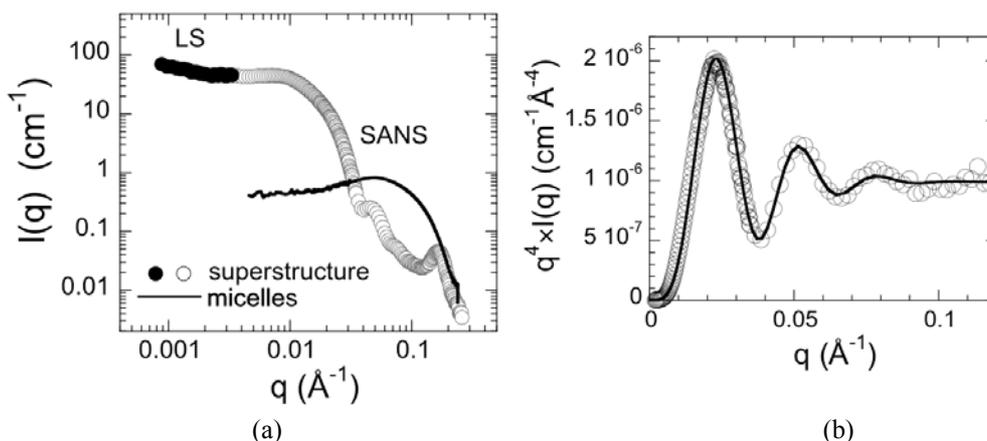

(a) (b)

**Figure 6 : (a)** Intensité diffusée (DNPA) par une solution de micelles ($Z = \infty$, c= 1 wt. %), ainsi que par une solution mixte (DNPA et diffusion de la lumière statique) de micelles et copolymère ($Z = 1$, c = 1 wt. %) dans le système DTAB/PANa$_{6.5k}$-b-PAM$_{37k}$. **(b)** Représentation de Porod, $q^4I(q)$ en fonction de q, pour la solution mixte, et comparaison à un modèle de sphère utilisant un diamètre de superstructures de 21.6 nm et une polydispersité de 0.16.

A ce stade de la discussion, deux commentaires s'imposent. Primo, à bien regarder la figure 6a, il y a une très faible contribution répulsive entre objets qui se manifeste par un léger pic aux petits angles, vers 0.01 Å$^{-1}$. En réalité, ce pic fausse notre extrapolation à q = 0, pour déterminer la masse de nos objets. Par contre, la position du pic peut être utilisée pour une deuxième estimation du nombre d'agrégation moyen : le déplacement de sa position dans le cas de micelles pures (0.05 Å$^{-1}$) vers 0.01 Å$^{-1}$ environ indique que la masse des objets en interaction a augmenté environ d'un facteur $(0.05/0.01)^3$, donc de un à environ une centaine de micelles par superstructure.

Secundo, on peut se demander où sont situées les micelles, puisque le domaine de Porod que nous avons trouvé correspond à la surface lisse des objets, et non des micelles. La réponse est que la Figure 6b se limite aux q intermédiaires, domaine où nous voyons effectivement la surface des gros objets. La diffusion aux grands angles, q > 0.1 Å$^{-1}$, nous réserve encore une surprise. En effet, on y trouve un pic situé à q = 0.16 Å$^{-1}$, suivi d'un deuxième régime de Porod. Notons que l'ajustement de la Figure 6b, si nous l'avions représenté sur toute la gamme de q, serait bien incapable de reproduire l'intensité piquée aux grands angles. Dans ce domaine, nous voyons effectivement la surface des micelles, comme on peut se convaincre par une simple estimation de la surface spécifique, ou en superposant la diffusion des micelles pures dans ce domaine. Et ces micelles semblent être en interaction forte, à juger par le pic prononcé. L'estimation de la distance typique centre-à-centre ($2\pi/q^* = 4$ nm) à partir de la position du pic q* nous révèle que les micelles doivent être localement très concentrées, les distances correspondant à des micelles au contact, i.e. deux rayons micellaires.

En résumé, nous avons déduit de quelques analyses plutôt simples que toutes les micelles s'organisent grâce aux copolymères sous forme de superstructures bien définies en taille (21.6 nm, 16% de polydispersité), contenant en moyenne $N_{agg}$= 100 - 250 micelles suivant la valeur de Z. Une vérification rapide permet d'ailleurs de voir que ces deux observations indépendantes - car correspondant à des mesures à des q différents - nous fournissent une information supplémentaire: la fraction volumique d'un tel assemblage vaut environ $N_{agg}(D_{mic}/D_{agg})^3 = 100*(3.6/21.6)^3 = 46$ %. Nous verrons plus tard que cet ordre de grandeur est cohérent avec la position du pic aux grands angles.

En confrontant le résultat de la DLS (un diamètre hydrodynamique de 70-80 nm) avec les résultats de la microscopie électronique et de la diffusion des neutrons ($D \approx 20$ nm), il apparaît que la DLS sonde une structure plus grande et moins dense que les deux autres techniques. Cela suggère que le tensioactif forme un cœur dense qui est observable en diffusion de neutrons et microscopie électronique, entouré d'une couche de polymère très hydraté. Naturellement, le PAM, probablement légèrement étiré, pourrait former la couronne, tandis que le tensioactif et la partie polyélectrolyte du copolymère formeraient le cœur du complexe.

*4.3.2 Les superstructures à base de nanoparticules vues par DNPA*

Nous avons fait le même type d'analyse par DNPA avec les superstructures formées dans les solutions mixtes de nanoparticules et de copolymère (PTEA$_{5K}$-*b*-PAM$_{11K}$) [32]. Les intensités diffusées par les suspensions de nanoparticules seules et par les mélanges sont superposées en Figure 7.

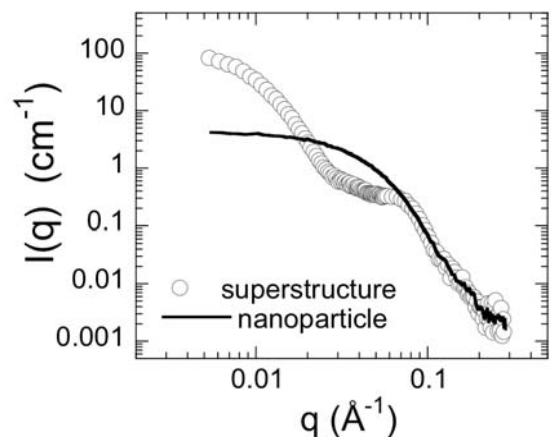

**Figure 7 :** Intensité diffusée (DNPA) par une solution de nanoparticules γ-Fe$_2$O$_3$-PAA$_{2K}$ (c = 1.38 wt. %), ainsi que par une solution mixte de nanoparticules et copolymère PTEA$_{5K}$-*b*-PAM$_{11K}$ (Z = 1, c= 1.7 wt. %).

Aux grands angles, nous observons un bon accord entre la structure des agrégats et des nanoparticules, ce qui suggère que les nanoparticules dominent le signal diffusé. Dans le cas des superstructures, l'intensité est nettement plus élevée aux petits angles, ce qui fournit une indication du nombre d'agrégation moyen suivant l'éq. (4), comme dans le cas des micelles. La courbure de l'intensité aux petits angles donne une idée de la dimension spatiale des superstructures, et il est possible d'obtenir un bon ajustement aux petits angles avec un modèle d'ellipsoïde de révolution[6, 32]. Comme avec les micelles, l'intensité présente un trou de corrélation aux moyens angles, avant de rejoindre l'intensité des nanoparticules seules, vers 0.07 Å$^{-1}$. Dans le cas des nanoparticules, comme pour les tensioactifs, l'intensité montre un pic d'interaction dans cette gamme de q, néanmoins d'intensité plus faible. Une relation simple permet d'ailleurs d'estimer grosso modo la taille des nanoparticules : si elles se touchent à l'intérieur des superstructures, la position du pic est donnée par $q^* = 2\pi/D$, ce qui donne D = 8.8 nm, légèrement plus élevé que le diamètre médian de la distribution des tailles (7.1 nm).

## 5. Modélisation

Nous avons vu dans la section précédente que la diffusion de neutrons aux petits angles permet de caractériser le cœur dense du complexe de colloïdes, qu'il soit à base de micelles ou de nanoparticules. Une comparaison au modèle simple de sphères ou ellipsoïdes polydisperses a permis d'estimer la géométrie du cœur, ainsi que la masse de micelles ou de nanoparticules contenues dans ce cœur. Si on souhaite accéder à sa structuration *interne*, il faut tenir compte de la diffusion par les particules en interaction dans le cœur. Pour des ensembles de particules monodisperses et de symétrie sphérique, l'intensité peut être factorisée en deux contributions [7]:

$$I(q) = I_0 \, S(q) \, F^2(q) \qquad (5)$$

Le facteur de forme $F^2(q)$ est mesuré à grande dilution, à condition d'avoir toujours les mêmes colloïdes (avec une concentration en tensioactifs supérieure à la cmc pour les micelles par exemple). Ici, nous avons choisi la notation $F^2(q)$ pour le facteur de forme normé à un aux petits angles. Le facteur de structure, $S(q)$, exprime les corrélations des centres des micelles ou nanoparticules, et il est donné par la transformée de Fourier de la fonction de corrélation de paires. Un petit calcul nous amène à la formule de Debye pour $S(q)$ [33] pour des ensembles moyennés sur toutes les orientations:

$$S(q) = 1 + \frac{2}{N_{agg}} \sum_{i>j}^{N_{agg}} \frac{\sin(qr_{ij})}{qr_{ij}} \qquad (6)$$

Ce facteur de structure peut être calculé pour n'importe quel ensemble de $N_{agg}$ objets sphériques, et la moyenne se fait sur beaucoup de configurations, afin d'obtenir une moyenne fiable et peu bruitée. Quelques propriétés de S(q) peuvent être déduites de l'équation (6) :

(a) Aux petits angles, $S(q) = N_{agg}$ pour des ensembles isolés de $N_{agg}$ nanoparticules.
(b) Aux grands angles, $S(q) = 1$.
(c) S(q) est piqué à une valeur de q qui correspond à la distance la plus probable, par exemple $q = 2\pi/D$ pour des particules de diamètre D au contact.

### 5.1 Modélisation par simulation Monte Carlo (directe)

Nous avons mis en place une modélisation Monte Carlo de la structure interne du cœur des superstructures formées par les micelles. Nous prenons pour acquis que le nombre de molécules de tensioactif par micelle n'a pas changé, puisque dans le régime de Porod, aux grands angles, l'intensité des superstructures se superpose avec le facteur de forme des micelles mesuré en solution diluée. Nous savons aussi que le nombre de micelles et la taille des superstructures, quantités extraites de l'intensité aux petits angles, correspondent à une densité compatible avec la position du pic à grands angles : en effet, l'évolution de la position du pic d'une solution de micelles dans l'eau extrapolée à 50% (en réalité, une transition de phase est observée) présenterait un pic situé vers 0.16 Å$^{-1}$, exactement comme observé dans nos superstructures de complexes de colloïdes.

Nous concluons que l'ajout de copolymère doublement hydrophile génère des complexes de colloïdes de taille connue (D de l'ordre de 25 nm, dépendant de Z), de masse connue (exprimée en nombre de micelles $N_{agg}$, de l'ordre de la centaine), et qu'il est constitué des mêmes micelles qu'en solution. Nous voudrions maintenant utiliser la formule de Debye, eq.(6), pour vérifier si un modèle de micelles empilées de manière dense dans le cœur de l'agrégat permet de reproduire les courbes expérimentales sur toute la gamme de q, ou s'il faut éventuellement tenir compte d'interactions spécifiques, e.g. dues aux copolymère. En résumé, nous devons calculer le facteur de forme d'un tel agrégat, et le comparer à l'intensité observée expérimentalement. Ceci est possible en deux étapes :

(1) Génération de configurations de $N_{agg}$ sphères (représentant les micelles) enfermées dans une sphère plus grande, de rayon $R_{coeur}$.

(2) Calcul de l'intensité diffusée par cet ensemble.

Le premier point paraît à priori simple ; il faut juste se méfier des corrélations introduites dans des ensembles trop denses. Pour éviter toute complication, nous avons généré un grand nombre de configurations en partant d'un ensemble peu dense ($R_{coeur}$ grand), et en diminuant pas à pas ce rayon tout en maintenant un mouvement Brownien des micelles à l'intérieur du cœur. Ce mécanisme est schématisé en Figure 8a.

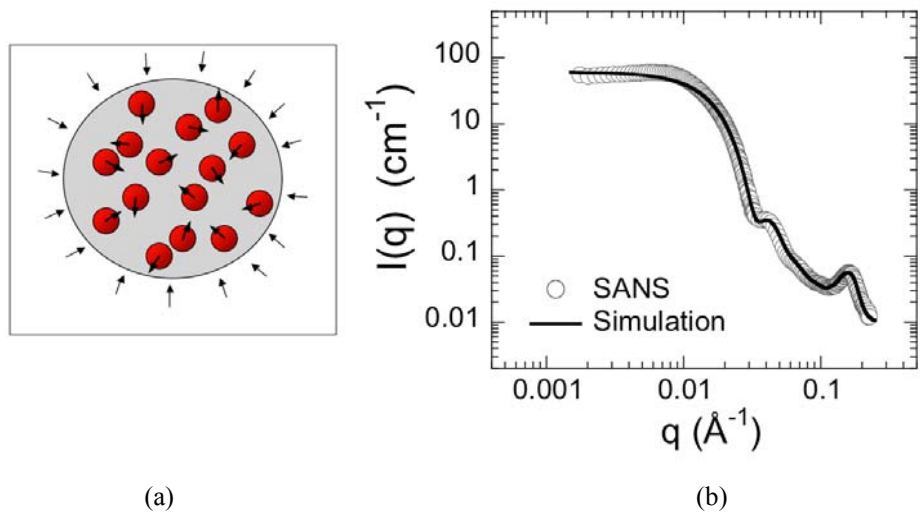

(a)                                                               (b)

**Figure 8 :** **(a)** Schéma de la simulation de Monte Carlo directe dans la phase de compaction des micelles dans le cœur de la superstructure. **(b)** Intensité diffusée (DNPA) par une solution mixte de micelles et copolymère (Z = 1, c = 1), dans le système DTAB/PANa$_{6.5k}$-*b*-PAM$_{37k}$, comparée à la prédiction de la simulation de Monte Carlo.

A ce stade, on peut aussi introduire une petite dose de polydispersité dans la taille des micelles, à condition que cela reste suffisamment faible pour ne pas introduire des corrélations. Nous voudrions par exemple rester loin du cas d'un placement préférentiel de très petites particules dans les interstices des grandes. Si l'ensemble est monodisperse (ou presque), on peut calculer l'intensité diffusée en utilisant la factorisation en facteur de forme et facteur de structure, eq. (5), et donc répondre au deuxième point ci-dessus. La validité de cette approximation est par ailleurs discutée dans la littérature, notamment par le groupe de R. Klein. En Figure 8b, nous montrons qu'il est effectivement possible de faire un tel calcul, et que le résultat est plutôt encourageant : nous reproduisons l'intensité globalement de manière satisfaisante, à l'exception de pic d'interaction entre superstructures, vers 0.01 Å$^{-1}$, qui n'est pas décrit par le modèle. Bien sur, nous avons travaillé en amont de l'analyse, ce qui fait que l'accord aux petits angles est juste par construction, tout autant que le domaine de Guinier des complexes. L'oscillation aux q intermédiaires est également bien reproduite, en générant une famille de configurations obéissant à la statistique déterminée par des fits simples montrées en section

4. Il est donc normal que ces « key features » soient bien décrits par notre modèle. Par contre, le modèle permet aussi d'ajuster les données aux grands angles, y compris le pic. Ceci indique que notre modèle est compatible avec l'expérience à toutes les échelles de longueur observées. Comme il est difficile d'inventer d'autres modèles répondant aux mêmes contraintes, celui-ci devient incontournable en pratique.

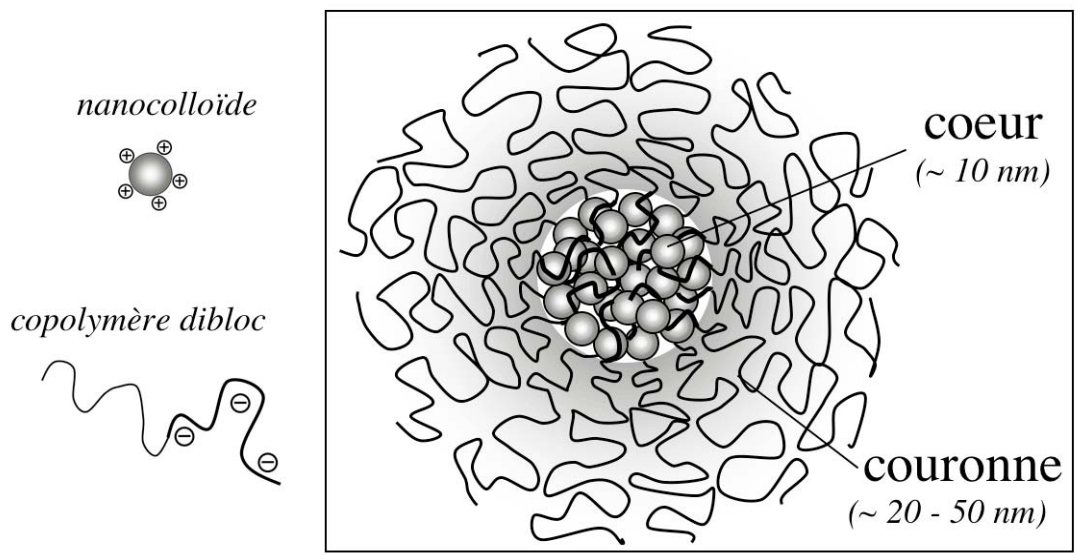

**Figure 9 :** Représentation schématique des superstructures à base de micelles et de copolymères doublement hydrophiles, chargé-neutres.

En résumé, nous avons obtenu une caractérisation structurale complète des superstructures formées par des micelles chargées et des copolymères doublement hydrophiles, chargé-neutres. Elles sont composées d'un cœur dense d'une centaine de micelles, en forte interaction. Ce cœur dense est observable en cryo-TEM et DNPA, et sa signature dans toute la gamme de q peut être décrite par un modèle de Monte Carlo simple. En diffusion de la lumière, où l'on sonde les interactions hydrodynamiques, la coquille très solvatée et peu dense contribue fortement au diamètre hydrodynamique, qui est donc notablement plus grand (70 – 80 nm) que le diamètre du cœur ($\approx$ 25 nm). Une représentation schématique est montrée en Figure 9.

## 5.2 Modélisation par simulation de Monte Carlo inverse (RMC)

Après avoir décrit les superstructures de micelles et de copolymères par une modélisation de Monte Carlo directe, nous nous intéressons maintenant aux complexes de colloïdes de nanoparticules. Nous avons modélisé leur structure par Monte Carlo inverse (reverse MC, ou RMC). Notons que la méthode est transposable à d'autres domaines [34]. L'analyse des courbes d'intensité, comme celle présentée en Figure 7, a montré qu'il s'agit probablement d'objets ellipsoïdaux [32]. Nous connaissons aussi la masse moyenne des superstructures. Une approche de Monte Carlo *directe* comme introduite en section 5.1 pour les complexes sphériques consiste à générer un ellipsoïde avec les paramètres trouvés ci-dessus, de le remplir de nanoparticules, pour déterminer S(q) grâce à l'équation (6). Ensuite, l'expérimentateur compare ce résultat aux données expérimentales pour valider le modèle. S'il ne possédait pas déjà des informations sur la géométrie moyenne des ellipsoïdes, il procéderait naturellement par itération, en variant ces paramètres 'à la main', afin de trouver le meilleur accord. Le résultat de cette méthode directe est double : en cas de réussite, nous déduisons les 'bons' paramètres comme l'excentricité, ou la masse moyenne des agrégats. De plus, dans notre pauvre monde de mesure d'intensités et non de phases et d'amplitudes, on se contenterait de dire que le modèle est *compatible* avec l'observation, i.e. que le modèle permet de reproduire les données. Souvent, il est difficile d'imaginer d'autres modèles 'physiques' ayant les mêmes qualités, et on finit par croire en la structure prédite. Si par contre l'expérimentateur ne parvient pas à trouver un jeu de paramètres reproduisant correctement l'observation, il est obligé d'écarter ce modèle définitivement. Il y a donc aussi de l'information dans l'échec.

Nous avons appliqué ici une autre méthode, dite de Monte Carlo *inverse*. Elle a été développée il y a quelques années pour décrire les fonctions de corrélations dans les liquides simples [35-37]. Elle a l'avantage de laisser totalement libre la géométrie des complexes, même si bien sûr il est toujours possible d'imposer des contraintes. Dans le cadre de cette modélisation, nous ne sommes donc plus esclaves d'une forme donnée; cependant, ce qui a été écrit ci-dessus sur les méthodes directes reste vrai : il faut d'abord avoir quelques idées assez minimalistes sur par exemple les interactions entre nanoparticules. L'algorithme proposé ci-dessous cherche ensuite un agencement compatible avec l'observation. S'il y parvient, nous avons trouvé une représentation possible de la réalité. Un échec par contre remet en question les contraintes minimalistes imposées au modèle.

Le principe de base est très simple pour des agrégats de taille finie, que nous supposons tous identiques dans un premier temps. Après un traitement primaire comme exposé en section 4, on connaît le nombre de particules dans un complexe, et on peut procéder par exemple de la manière suivante :

(a) Génération d'une configuration initiale compatible avec les contraintes physiques, comme par exemple le non recouvrement entre particules, ou une dimension maximale des agrégats.
(b) Détermination du facteur de structure S(q) par la formule de Debye, en utilisant l'éq. (6).

(c) Exécution d'un pas de Monte Carlo pour faire évoluer la configuration initiale. Ce pas est illustré en Figure 10a. Il doit être compatible avec les contraintes physiques, et si possible permettre une rapide exploration de l'espace des phases.

(d) Calcul du nouveau facteur de structure.

(e) Comparaison de l'accord entre expérience et modélisation, avant et après le pas Monte Carlo. L'idée est d'accepter le pas s'il a amélioré l'accord, et le refuser sinon. Ici il est possible (et recommandé) d'introduire des méthodes de recuit simulé (« simulated annealing »), ce qui améliore en général la convergence de l'algorithme [38].

En mots simples, une simulation de RMC correspond à la construction d'un modèle relativement complexe d'un agrégat, car fait avec une sous-structure, la nanoparticule. Ce modèle possède un grand nombre de paramètres, qui sont grosso modo les degrés de liberté de mouvement des particules. Comme il est impossible d'optimiser le fit en variant un à un tous ces paramètres car tout est lié, le mécanisme de Monte Carlo permet d'explorer l'espace des phases de manière aléatoire.

A ce point, on pourrait croire qu'un fit automatique avec un si grand nombre de paramètres convergerait forcément, après tout il est bien connu qu'avec cinq paramètres on peut fitter un éléphant. Nous avons montré dans le passé que ce n'est souvent pas le cas, notamment à cause des contraintes physiques imposées via les règles de la simulation. Un certain savoir-faire reste nécessaire, non seulement pour définir un jeu de règles 'physiques', mais aussi pour trouver des pas de MC qui explorent l'espace des phases de manière efficace, afin de trouver le meilleur accord possible.

L'exemple des agrégats de nanoparticules magnétiques est illustré en Figure 10b [32]. Comme le facteur de forme des nanoparticules est connu - car mesurable facilement à grande dilution -, nous avons choisi de travailler sur S(q) plutôt que sur I(q), les deux fonctions étant liées par l'éq.(5). Dans ce cas de figure, les caractéristiques de S(q) sont relativement peu marquées, et il est avantageux d'analyser S(q) directement. La configuration initiale peut être quelconque, par exemple un cristallite cubique du bon nombre de nanoparticules. Imposer ce dernier paramètre par la masse moyenne mesurée aux petits angles est un exemple de contrainte physique diminuant considérablement l'espace des phases à explorer. En quelques pas de Monte Carlo inverse par particule, le cristallite 'fond' et se réorganise sous une forme ellipsoïdale, de structure et densité interne données. Rappelons que la forme ellipsoïdale n'est pas imposée par l'algorithme. En revanche, elle est compatible avec les résultats de cryo-TEM (Figure 5), ainsi qu'avec les ajustements globaux de l'intensité aux petits angles. Comme on peut remarquer en Figure 10b, le facteur de structure de la configuration résultante reproduit plusieurs caractéristiques du S(q) expérimental : la forte diminution aux petits angles, le trou de corrélation, ainsi que le faible maximum correspondant aux corrélations entre particules voisines.

Lors de la simulation RMC, l'accord entre modèle et observation, quantifié par $\chi^2$, la somme des écarts quadratiques, diminue très vite en cas de modèle adapté, pour ensuite fluctuer autour d'un plateau [34]. Les détails de la fluctuation dépendent de l'implémentation de l'algorithme. Ils expriment le fait que nous avons trouvé un ensemble de configurations en accord avec l'expérience ($\chi^2$ acceptable). Chaque déplacement supplémentaire le long de ce plateau génère une nouvelle configuration, dont les caractéristiques globales restent inchangées. Comme ces caractéristiques sont la traduction directe de leur signature dans l'espace réciproque (décroissance globale de l'intensité avec q, position et profondeur du trou de corrélation), elles sont robustes et crédibles. Dans le cas des nanoparticules, il a ainsi été possible d'estimer l'épaisseur de la couche en surface des nanoparticules ($PAA_{2k}$ par exemple), par les fits RMC. En résumé, l'algorithme de Monte Carlo inverse a plusieurs mérites : (a) il permet extraire les caractéristiques principales dans le cadre d'un modèle à minima. (b) il produit des représentations dans l'espace direct qui sont compatibles avec l'expérience. Et (c) on peut ensuite utiliser ces représentations pour déterminer des fonctions de corrélation. L'algorithme RMC est donc une méthode stochastique de solution du problème inverse, comme la 'indirect Fourier transform' (IFT) [39], mais soumis à des règles physiques.

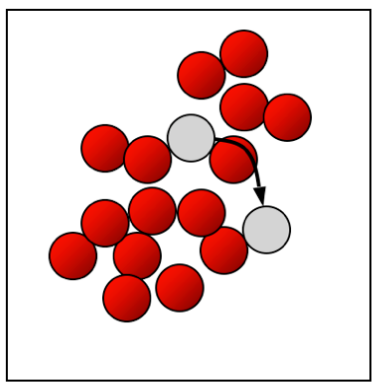
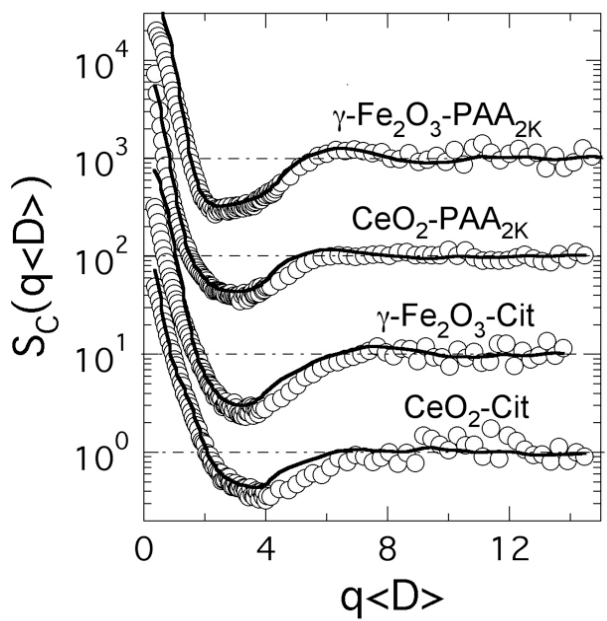

(a)            (b)

**Figure 10 :** **(a)** Schéma de la simulation de Monte Carlo inverse (RMC) décrivant le cœur de la superstructure. **(b)** Intensité diffusée (DNPA) par une solution mixte de plusieurs types de nanoparticules et copolymère (PTEA-*b*-PAM), comparée à la prédiction de la simulation de Monte Carlo inverse [32]. <D> dénote le diamètre moyen des nanoparticules, cette renormalisation permettant de comparer des NPs de taille différente.

Dans la littérature, il y a eu des discussions sur l'unicité des solutions. Comme nous avons essayé de le dire ci-dessus, ce débat semble vain : par définition, l'algorithme produit un ensemble de solutions, toutes compatibles avec l'observation. Etant donné l'absence d'information de phase, le résultat est donc aussi peu 'unique' que toutes les autres modélisations. Il nous reste cependant un dernier point, la polydispersité. La RMC telle que nous l'avons décrite représente un agrégat moyen. Il est a priori facile d'introduire une polydispersité dans l'algorithme, mais si l'accord est déjà très bon pour un agrégat moyen, ce nouveau paramètre n'apportera pas d'informations supplémentaires, sauf dans le cas ou la seule polydispersité acceptable serait … le cas monodisperse.

## 6. Conclusion

Dans ce chapitre, nous avons discuté la formation de complexes de colloïdes. La structure de ces objets supra-colloïdaux a été étudiée en combinant plusieurs techniques expérimentales : deux observations 'statiques' dans l'espace directe (cryo-TEM) et réciproque (DNPA), et une technique basée sur les propriétés de diffusion, la diffusion dynamique de la lumière (DLS). Les échelles sondées vont environ du nanomètre à la centaine de nanomètres. Nous avons essayé de décrire le concept, l'apport, ainsi que les limites de chaque technique.

Les complexes de colloïdes se forment spontanément en solution aqueuse, en mélangeant des petites particules chargées (micelles ou nanoparticules) avec des copolymères à bloc doublement hydrophiles, chargé-neutres, le bloc chargé étant de charge opposée par rapport aux particules. Le mécanisme de formation probable est montré schématiquement en Figure 11. Les particules colloïdales chargées sont pontées par les blocs chargés du copolymère. Ce dernier induit donc une attraction effective entre les particules. Les blocs hydrophiles se trouvent expulsés du complexe, et forment – au fur et à mesure que le colloïde croît – une couche de plus en plus dense de polymère hydraté, la couronne. A partir d'une certaine taille, cette couche stabilise le complexe stériquement, et l'empêche de grandir davantage. Ce mécanisme expliquerait la structure finale – cœur et couronne -, ainsi que la monodispersité des objets, comme observée en diffusion et microscopie électronique.

24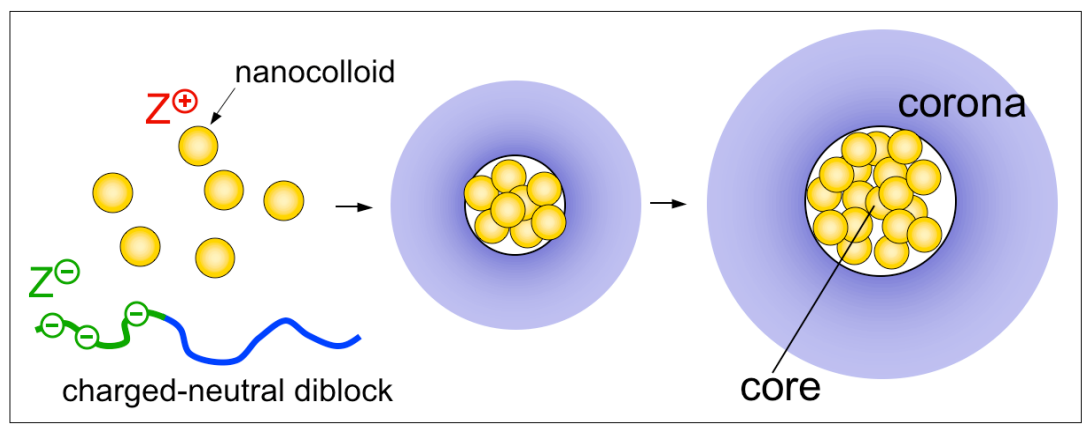

**Figure 11 :** Mécanisme de formation de complexes de colloïdes par des colloïdes en interaction attractive induite par un copolymère chargé-neutre.

La caractérisation structurale des complexes de colloïdes, basée en partie sur des simulations numériques pour interpréter les résultats de diffusion aux petits angles, nous a permis de suivre l'impact de chaque modification des briques élémentaires – telles que la taille, la nature de l'enrobage et de la charge – sur la formation des suprastructures. La stabilité accrue des nanostructures à base de tensioactifs ou de nanoparticules est attribuée au fait que le processus d'association est un phénomène hors équilibre. Une fois que les nanostructures sont formées, elles restent inchangées sur des périodes pouvant aller jusqu'à plusieurs années. Cela est dû aux fortes interactions électrostatiques entre les constituants de départ. Parce que la stratégie présente est simple et versatile, elle devrait ouvrir des perspectives intéressantes pour le développement de colloïdes multimodaux activables à distance, des nano-outils pour la biophysique et la biomédecine tels que des sondes, des senseurs ou des éléments pour la microfluidique, tels que des agitateurs. Il a été montré par exemple que les agrégats de nanoparticules d'oxyde de fer avaient un meilleur contraste en imagerie par résonance magnétique que les particules individuelles, ouvrant ainsi la voie à leur utilisation en biomédecine pour le diagnostique de déficiences hépatiques.

**Remerciements**

Nous tenons à remercier nos collègues qui ont participé à ce travail et sans qui cette recherche n'aurait pas pu voir le jour. C'est avec plaisir que nous remercions Jean-Christophe Castaing, Jean-Paul Chapel, Galder Cristobal, Jérôme Fresnais, Pascal Hervé, Eléna Ishow, Mikel Morvan, Régine Perzynski, Ling Qi, Olivier Sandre, Amit Sehgal, Minhao Yan et Kazuhiko Yokota. Notre gratitude va aussi à Bernard Cabane, Valérie Cabuil, Andrejs Cebers, Martien Cohen-Stuart, Mathias Destarac, Eric Kaler, Christophe Lavelle, Sébastien Lecommandoux, Patrick Maestro, Nathalie Mignet, Lennard Picullel, Didier Roux, Serge Stoll pour de nombreuses et passionnantes discussions. Le Laboratoire Léon Brillouin à Saclay (Annie Brûlet, Fabrice Cousin) et l'Institut Laue-Langevin à Grenoble (Isabelle Grillo, Ralf Schweins) sont remerciés pour leur soutien technique et financier. Il en va de

**Références**

[1] Pankhurst, Q.; Connolly, J.; Jones, S.; Dobson, J., Applications of Magnetic Nanoparticles in Biomedicine. *J Phys D: Appl Phys.* **2003,** 36, R167 – R81.

[2] Michalet, X.; Pinaud, F.; Bentolila, L.; Tsay, J.; Doose, S.; Li, J.; Sundaresan, G.; Wu, A.; Gambhir, S.; Weiss, S., Quantum Dots for Live Cells, in Vivo Imaging, and Diagnostics. *Science* **2005,** 307, 538 - 44.

[3] Gupta, A.; Gupta, M., Synthesis and surface engineering of iron oxide nanoparticles for biomedical applications. *Biomaterials* **2005,** 26, (18), 3995-4021.

[4] Jain, P.; Huang, X.; El-Sayed, I.; El-Sayed, M., Noble Metals on the Nanoscale: Optical and Photothermal Properties and Some Applications in Imaging, Sensing, Biology, and Medicine. *Accounts of Chemical Research* **2008,** 41, (12), 1578-86.

[5] Naka, K.; Chujo, Y., Nanohybridized Synthesis of Metal Nanoparticles and Their Organization. In *Nanohybridization of Organic-Inorganic Materials*, 2009; pp 3-40.

[6] Lindner, P., *Neutrons, X-ray and Light Scattering*. North Holland, Elsevier: 2002.

[7] Belloni, L., La Diffusion de Neutrons aux Petits Angles: mécanique statistique des liquides et traitement des données. *Ce volume*.

[8] Berret, J.-F., Evidence of overcharging in the complexation between oppositely charged polymers and surfactants. *J. Chem. Phys.* **2005,** 123, (16), 164703.

[9] Destarac, M.; Bzducha, W.; Taton, D.; Gauthier-Gillaizeau, I.; Zard, S., Xanthates as chain-transfer agents in controlled radical polymerization (MADIX): Structural effect of the O-alkyl group. *MACROMOLECULAR RAPID COMMUNICATIONS* **2002,** 23, (17), 1049-1054.

[10] McGrath, K., Phase Behavior of Dodecyltrymethylammonium Bromide/Water Mixtures. *Langmuir* **1995,** 11, 1835 - 1839.

[11] Konop, A.; Colby, R., Role of Condensed Counterions in the Thermodynamics of Surfactant Micelle Formation with and without Oppositely Charged Polyelectrolytes. *Langmuir* **1999,** 15, 58 - 65.

[12] Bee, A.; Massart, R.; Neveu, S., Synthesis of very fine maghemite particles. *J Magn Magn Mat* **1995,** 149, 6-9.

[13] Massart, R.; Dubois, E.; Cabuil, V.; Hasmonay, E., Preparation and properties of monodisperse magnetic fluids. *J Magn Magn Mat* **1995,** 149, 1 - 5.

[14] Berret, J.-F.; Sandre, O.; Mauger, A., Size distribution of superparamagnetic particles determined by magnetic sedimentation. *Langmuir* **2007,** 23, (6), 2993-2999.

[15] Dubois, E.; Cabuil, V.; Boue, F.; Perzynski, R., Structural analogy between aqueous and oily magnetic fluids. *J Chem Phys* **1999,** 111, 7147 - 7160.

[16] Lucas, I.; Durand-Vidal, S.; Dubois, E.; Chevalet, J.; Turq, P., Surface Charge Density of Maghemite Nanoparticles: Role of Electrostatics in the Proton Exchange. *The Journal of Physical Chemistry C* **2007,** 111, 18568-76.

[17] Berret, J.-F., Stoichiometry of electrostatic complexes determined by light scattering. *Macromolecules* **2007,** 40, (12), 4260-4266.

[18] Berret, J.-F.; Sehgal, A.; Morvan, M.; Sandre, O.; Vacher, A.; Airiau, M., Stable oxide nanoparticle clusters obtained by complexation. *J. Colloid Interface Sci.* **2006,** 303, (1), 315-318.
[19] Chanteau, B.; Fresnais, J.; Berret, J. F., Electrosteric Enhanced Stability of Functional Sub-10 nm Cerium and Iron Oxide Particles in Cell Culture Medium. *Langmuir* **2009,** 25, (16), 9064-9070.
[20] Fresnais, J.; Berret, J.-F.; Frka-Petesic, B.; Sandre, O.; Perzynski, R., Electrostatic Co-Assembly of Iron Oxide Nanoparticles and Polymers: Towards the Generation of Highly Persistent Superparamagnetic Nanorods. *Adv. Mater.* **2008,** 20, (20), 3877-3881.
[21] Fresnais, J.; Lavelle, C.; Berret, J. F., Nanoparticle Aggregation Controlled by Desalting Kinetics. *The Journal of Physical Chemistry C* **2009,** 113, (37), 16371-16379.
[22] Yan, M.; Fresnais, J.; Berret, J. F., Growth mechanism of nanostructured superparamagnetic rods obtained by electrostatic co-assembly. *Soft Matter* **2010,** 6, (9), 1997-2005.
[23] Berret, J.-F.; Vigolo, B.; Eng, R.; Herve, P.; Grillo, I.; Yang, L., Electrostatic self-assembly of oppositely charged copolymers and surfactants: A light, neutron, and X-ray scattering study. *Macromolecules* **2004,** 37, (13), 4922-4930.
[24] Ashbaugh, H.; Lindman, B., Swelling and Structural Changes of Oppositely Charged Polyelectrolyte Gel-Mixed Surfactant Complexes. *Macromolecules* **2001,** 34, 1522-25.
[25] Ilekti, P.; Piculell, L.; Tournilhac, F.; Cabane, B., How to Concentrate an Aqueous Polyelectrolyte/Surfactant Mixtures by Adding Water. *J Phys Chem B.* **1998,** 102, 344 - 51.
[26] Kogej, K.; Evmenenko, G.; Theunissen, E.; Berghmans, H.; Reynaers, H., Investigation of Structures in Polyelectrolytre/Surfactant Complexes by X-ray Scattering. *Langmuir* **2001,** 17, 3175-84.
[27] Hansson, P., Self-Assembly of Ionic Surfactants in Polyelectrolyte Solutions : A Model for Mixtures of Opposite Charge. *Langmuir* **2001,** 17, 4167-79.
[28] Nallet, F., Introduction à la diffusion aux petits angles. *Ce volume*.
[29] Berret, J.-F.; Herve, P.; Aguerre-Chariol, O.; Oberdisse, J., Colloidal complexes obtained from charged block copolymers and surfactants: A comparison between small-angle neutron scattering, Cryo-TEM, and simulations. *J. Phys. Chem. B* **2003,** 107, (32), 8111-8118.
[30] Klucker, R.; Schosseler, F., Small-Angle Neutron and X-ray Scattering Comparative Study of Polyacrylamide and Hydrophobically Modified Polyacrylamide in Aqueous Solution. *Macromolecules,* **1997,** 30, (14), 4228-4231.
[31] Belloni, L., Ionic Condensation and Charge Renormalization in Colloid Suspensions. *Colloids Surf A.* **1998,** 140, 227 - 243.
[32] Fresnais , J.; Berret, J.-F.; Qi, L.; Chapel, J.-P.; Castaing, J.-C.; Sandre, O.; Frka-Petesic, B.; Perzynski, R.; Oberdisse, J.; Cousin, F., Universal scattering behavior of coassembled nanoparticle-polymer clusters. *Phys Rev E* **2008**, 78, 040401(R).
[33] Debye, P., *Phys. Coll. Chem.* **1947,** 51, 18.
[34] Oberdisse, J.; Hine, P.; Pyckhout-Hintzen, W., Structure of interacting aggregates of silica particles for elastomer reinforcement. *Soft Matter* **2007,** 2, 476-485.
[35] McGreevy, R. L., Reverse Monte Carlo modelling. *Journal of Physics: Condensed Matter* **2001,** 13, R877-R913.
[36] McGreevy, R. L.; Zetterstrom, P., To RMC or not to RMC? The use of reverse Monte Carlo modelling. *Current Opinion in Solid State & Materials Science* **2003,** 7, (1), 41-47.
[37] Pusztai, L.; Dominguez, H.; Pizio, O. A., Reverse Monte Carlo modeling of the structure of colloidal aggregates. *Journal of Colloid and Interface Science* **2004,** 277, (2), 327-334.
[38] Press, W. H.; Teukolsky, S. A.; Vetterling, W. T.; Flannery, B. P., *Numerical recipes in Fortran*. 2nd ed.; Cambridge University Press: Cambridge, 1992.
[39] Glatter, O., *J. Appl. Cryst.* **1977,** 10, 415-421.